\documentclass[a4paper,fleqn,usenatbib,useAMS]{mnras}

\usepackage{graphicx}	% Including figure files
\usepackage{amsmath}	% Advanced maths commands
\usepackage{amssymb}	% Extra maths symbols
\usepackage{multicol}        % Multi-column entries in tables
\usepackage{bm}		% Bold maths symbols, including upright Greek
\usepackage{pdflscape}	% Landscape pages
\newcommand{\AustralianJournalOfPhysics}{Aust. J. Phys.}

\newcommand{\CanadianJournalOfPhysics}{Can. J. Phys.}

\newcommand{\ChemicalReviews}{Chem. Rev.}

\newcommand{\JournalOfPhysicsBAtomicMolecularOpticalPhysics}{J. Phys. B: At. Mol. Opt. Phys.} %1988 and onwards

\newcommand{\JournalOfPhysicalChemicalReferenceData}{J. Phys. Chem. Ref. Data}
\newcommand{\JournalOfChemicalPhysics}{J. Chem. Phys.}
\newcommand{\PhysicalReview}{Phys. Rev.}
\newcommand{\PlanetarySpaceScience}{Planet. Space Sci.}
\newcommand{\MolecularPhysics}{Mol. Phys.}
\newcommand{\ReportsOnProgressInPhysics}{Rep. Prog. Phys.}

\usepackage[T1]{fontenc}
\usepackage{ae,aecompl}
\usepackage{newtxtext,newtxmath}

\title[C-type shock modelling]{C-type shock modelling -- the effect of new H$_2$--H collisional rate coefficients}

\author[A. V. Nesterenok et al.] {A.V. Nesterenok$^{1}$\thanks{E-mail:alex-n10@yandex.ru},
D. Bossion$^2$, Y. Scribano$^2$ and F. Lique$^{3}$ \\
$^{1}$ Ioffe Institute, 26 Polytechnicheskaya St., 194021 Saint Petersburg, Russia \\
$^{2}$ LUPM-UMR 5299, CNRS-Universit\'e de Montpellier, Place E. Bataillon, F-34095 Montpellier, France \\
$^{3}$ LOMC-UMR 6294, CNRS-Universit\'e du Havre, 25~rue Philippe Lebon, BP~1123, F-76063 Le Havre, France}

\date{Accepted XXX. Received YYY; in original form ZZZ}
\pubyear{2015}

\begin{document}
\label{firstpage}
\pagerange{\pageref{firstpage}--\pageref{lastpage}}
\maketitle

\begin{abstract}
We consider collisional excitation of H$_2$ molecules in C-type shocks propagating in dense molecular clouds. New data on collisional rate coefficients for (de-)excitation of H$_2$ molecule in collisions with H atoms and new H$_2$ dissociation rates are used. The new H$_2$--H collisional data are state of the art and are based on the most accurate H$_3$ potential energy surface. We re-examine the excitation of rotational levels of H$_2$ molecule, the para-to-ortho-H$_2$ conversion, and H$_2$ dissociation by H$_2$--H collisions. At cosmic ray ionization rates $\zeta \geq 10^{-16}$~s$^{-1}$ and at moderate shock speeds, the H/H$_2$ ratio at the shock front is mainly determined by the cosmic ray ionization rate. The H$_2$--H collisions play the main role in the para-to-ortho-H$_2$ conversion and, at $\zeta \geq 10^{-15}$~s$^{-1}$, in the excitation of vibrationally excited states of H$_2$ molecule in the shock. The H$_2$ ortho-to-para ratio (OPR) of the shocked gas and column densities of rotational levels of vibrationally excited states of H$_2$ are found to depend strongly on the cosmic ray ionization rate. We discuss the applicability of the presented results to interpretation of observations of H$_2$ emission in supernova remnants.
\end{abstract}


\section{Introduction}
The hydrogen molecule has two possible nuclear spin states due to the proton spin of $1/2$ -- ortho-H$_2$ and para-H$_2$. In the electronic ground state, the rotational levels of ortho-H$_2$ have odd values of the angular momentum $j$ while the levels of para-H$_2$ have even $j$ values. During radiative processes and non-reactive collisions of H$_2$ with other atoms and molecules, only transitions with even values of angular momentum change $\Delta j$ are permitted, thus preserving the ortho-to-para ratio (OPR). The ortho-/para-H$_2$ interconversion in the interstellar gas is possible via reactive collisions (which include exchange of protons) of H$_2$ with H, H$^+$, H$_3^+$ and other species, as well as via chemical reactions (H$_2$ formation on dust grains). The conversion between ortho-H$_2$ and para-H$_2$ also takes place on the surface of interstellar dust grains \citep{Bron2016,Bovino2017,Furuya2019}.

The H$_2$ OPR is an important parameter of the cold interstellar medium -- the H$_2$ OPR is a controller of the cold cloud deuteration chemistry (e.g., \citealt{PineaudesForets1991,Flower2006,Pagani2009}), it affects the molecular excitation \citep{Troscompt2009}, determines the heat capacity of the gas \citep{Vaytet2014}. In the cold molecular gas, the H$_2$ OPR slowly decays to a small value -- 0.001 or even less \citep{Pagani2011}. The ortho-H$_2$ destruction is compensated by the H$_2$ forming on dust grains and by destruction of ions \citep{Flower2006,LeBourlot1991}. The OPR of H$_2$ may not be at thermal equilibrium because the time-scale of the ortho-to-para conversion can be significantly longer than the time-scale of dynamic evolution of the molecular gas \citep{Flower2006}. Reactive collisions of H$_2$ molecule with hydrogen atoms have a substantial activation energy, $ \simeq 5000$ K \citep{Lique2014}. This channel becomes dominant in a hot molecular gas, for example, behind interstellar shocks. 

Shock waves are born due to large pressure disturbances in the interstellar medium caused by star-driven jets and winds, supernova explosions, and collisions between molecular clouds. Interstellar shock waves can be distinguished depending on the intensity of magnetic field, the ionization fraction of the gas, and the shock speed \citep{Flower2007,Draine2011}. Here we focus on magnetohydrodynamic non-dissociative shock waves -- C-type shocks. Molecular hydrogen is an important tracer of non-dissociative shocks as it dominates cooling of the shocked gas (e.g., \citealt{Kaufman1996,Flower2015,Tram2018}). Excitation of molecular hydrogen and evolution of the H$_2$ OPR in interstellar shocks were studied theoretically by \citet{Timmermann1998,Wilgenbus2000}, and observed in many Galactic sources (e.g., \citealt{Neufeld2006,Neufeld2007,Shinn2011,Shinn2012,Neufeld2019}). The main parameters which determine the speed of para-to-ortho-H$_2$ conversion in a C-type shock are the fraction of atomic hydrogen and the gas temperature. The rate coefficients for collisions involving H$_2$ molecules and H atoms are crucial for modelling of H$_2$ OPR in such shocks.  
% Kaufman1996,

The interstellar ultraviolet (UV) radiation field cannot penetrate into the interiors of dark clouds, and thus, cosmic ray particles and X-ray radiation (if present) appear the main drivers of the gas phase chemistry \citep{Larsson2012}. The abundance of atomic H in dark clouds is determined by the interplay between H$_2$ destruction by cosmic rays and its formation on dust grains \citep{Goldsmith2005,Padovani2018}. At high ionization rates ($\zeta \gtrsim 10^{-15}$~s$^{-1}$), the atomic hydrogen can have considerable abundance of the order of $10^{-3} - 10^{-2}$ or even higher. Once a shock passes through molecular gas at such physical conditions, H$_2$--H collisions become important not only for the ortho-/para-state exchange of H$_2$ but also for the ro-vibrational excitation of H$_2$ molecules (e.g., \citealt{Neufeld2008}). In relatively fast C-type shocks, the abundance of H atoms can be higher due to partial dissociation of H$_2$ molecules.

\citet{Lique2015} has performed nearly exact quantum time-independent calculations of the rate coefficients for the collisional (de-)excitation of H$_2$ by H atoms. These new data are based on a highly accurate H$_3$ global potential energy surface, and the reactive hydrogen exchange channels are taken into account rigorously. \citet{Bossion2018} have performed quasi-classical trajectory (QCT) calculations of rate coefficients for the collisional (de-)excitation of H$_2$ by H atoms (including the three-body collisional dissociation) for almost all rotational energy levels of the ground electronic state of H$_2$ molecule. In the present paper, the new rate coefficients of H$_2$--H collisions are incorporated into the model of C-type shock propagating in a dense molecular cloud \citep{Nesterenok2018}. We re-examine (i) the excitation of rotational levels of H$_2$ molecule in C-type shocks, (ii) the para-to-ortho-H$_2$ conversion, and (iii) the H$_2$ dissociation by H$_2$--H collisions. The effect of the elevated levels of cosmic ray ionization rate on these processes is also studied.

% F. Lique: You may emphasis a bit more the (main) differences between previous and new collisional data. This would help the discussion of the results. Also, you can clearly state that the new data are state of the art and more accurate than the previous one because based on the most accurate H3 potential energy surface and because of fully quantum without approximation for Lique et al. 2015.

\section{Description of the calculations}
\subsection{H$_2$ collisional rate coefficients}
We take into account 298 rotational levels of the ground electronic state of H$_2$ molecule for which the Einstein coefficients are given by \citet{Wolniewicz1998}. The level energies of H$_2$ are taken from \citet{Dabrowski1984}.

The data on collisional (de-)excitation of H$_2$ by H atoms by \citet{Wrathmall2007}, which are usually used in astrophysical modelling, are restricted to non-reactive scattering. The treatment of ortho-/para-H$_2$ interconversion was usually adopted according to \citet{LeBourlot1999}, that in turn is based on the results of QCT calculations by \citet{Mandy1993,Martin1995} and laboratory experiments \citep{Schulz1965,Schofield1967}. Until recently, the contribution of the reactive scattering channels to the H$_2$ excitation by H atoms remained a significant source of uncertainty, see discussion by \citet{Wrathmall2007,Lique2014}. 

\citet{Lique2015} reported time-independent quantum mechanical calculations of rate coefficients for the collisional \mbox{(de-)excitation} of H$_2$ by H atoms. The calculations were based on a high accuracy global potential energy surface (PES) by \citet{Mielke2002}. These new data are the first computed for the ortho-/para-H$_2$ interconversion at high temperature and using a purely quantum approach. The data were obtained for the lowest 54 rotational levels of H$_2$ with internal energies up to 22~000~K (highest energy level is $v = 3$, $j = 8$) and for kinetic temperatures ranging from 100 to 5000~K. 

\citet{Bossion2018} presented QCT calculations of rate coefficients for the ro-vibrational (de-)excitation of H$_2$ by H atoms for temperatures up to 15~000~K, and for almost all rotational energy levels of the ground electronic state of H$_2$ molecule. Their calculations are based on the same PES by \citet{Mielke2002}. The ortho-/para-H$_2$ interconversion and the three-body collisional dissociation are included in their calculations. For this study, \citet{Bossion2018} calculations have been extended to all the bound states of H$_2$ molecule (internal energy up to dissociation limit $\approx 55~100$~K) and up to 20~000~K of collisional energy for the rate constant (the maximal temperature attained by the neutral gas in the high speed shocks in our simulations). For this high temperature regime we extended the QCT cross sections up to 100~000~K of collisional energy in order to ensure convergence on the rate constants. We used an energy-step of 2000~K up to that limit. Those QCT calculations were performed considering only the ground electronic state of H$_3$. Still the non-adiabatic effects are not expected to strongly impact the results (the first excited electronic state lies for internal energies over 58 000 K), and integrating the cross sections over the whole energy domain induces an averaging effect. Moreover \citet{Bouakline2010} studied the geometric-phase effect on the state-to-state cross section of the H$_3$ system and found little contribution of the non-adiabatic effects on the dynamics. It is to be noted that we considered the quasi-bound states as pertaining to the three-body dissociation channel. This assertion remains valid as long as the average lifetime of the quasi-bound molecules is lower than the typical collision time, this is true for low to moderate density media.

For H$_2$--H$_2$ collisions, the data by \citet{Wan2018} were used in our calculations. \citet{Wan2018} carried out quantum mechanical close-coupling calculations of collisional (de-)excitation rates based on the H$_2$--H$_2$ PES developed by \citet{Patkowski2008}. Initial rotational levels $j = 2-31$ of the ground vibrational state and kinetic temperatures up to 10~000~K were considered in their calculations. For the transitions involving other energy levels, the data by \citet{Flower1998b,Flower1999} are used in our simulations. We take into account the ro-vibrational excitation of H$_2$ molecule by fast electrons produced by cosmic ray ionization of the molecular gas \citep{Tine1997}. \citet{Gonzalez-Lezana2017} studied ortho-/para-H$_2$ conversion in collisions with H$^+$ and provided collisional rate coefficients for H$_2$ transitions between lowest rotational levels belonging to $v=0-3$ vibrational states. We use their data in our simulations. Table~\ref{table:h2collcoeff} presents data on collisional rate coefficients for H$_2$ molecule used in our calculations.

The dissociation of H$_2$ molecule by H$_2$ impact is taken into account \citep{Martin1998}. \cite{Martin1998} considered H$_2$ dissociation from the ground vibration--rotation state. The rate of dissociation of vibrationally excited H$_2$ molecule by H$_2$ impact is estimated using dissociation coefficients given by \citet{Ceballos2002}.

\begin{table}
\caption{Data on collisional rate coefficients for H$_2$ molecule. \label{table:h2collcoeff}}
\begin{tabular}{ll}
\hline
Collisional partner & Reference \\
\hline
H$_2$ & \citet{Wan2018}, \\
& \citet{Flower1998b}, \citet{Flower1999} \\
He & \citet{Flower1998a} \\
H & \citet{Lique2015}, \citet{Bossion2018} \\ 
thermal e$^{-}$ & \citet{Gerjuoy1955}, \citet{Ehrhardt1968}, \\
 & \citet{England1988}, \citet{Yoon2008} \\ 
non-thermal e$^{-}$ & \citet{Tine1997} \\
H$^+$ & \citet{Gonzalez-Lezana2017} \\
\hline
\end{tabular}
\end{table}

\subsection{C-type shock model}
The simulations of C-type shock consist of two steps: (i) modelling of the chemical and thermal evolution of a static molecular cloud and (ii) the shock simulations. A complete gas--grain chemical network is taken into account --  the gas-phase chemistry, the adsorption of gas species on dust grains and several desorption mechanisms, the chemical reactions on the surface of dust grains, the ion neutralization on dust grains. The population densities of energy levels of ions \ion{C}{i}, \ion{C}{ii} and \ion{O}{i} and molecules H$_2$, CO, H$_2$O are computed in parallel with the dynamical and chemical rate equations. The detailed information on the calculations can be found in the paper by \citet{Nesterenok2018}. In the appendix \ref{app_chemistry} we give brief overview of the model updates. 

Table~\ref{table:modelparam} summarizes the range of parameters of the shock wave explored in simulations. We assume that molecular cloud is shielded from the interstellar radiation field. The ionization rate $\zeta$ of the molecular gas is assumed to be due to cosmic rays, but we notice that X-rays affect molecular gas similarly \citep{Mackey2019}. The cosmic ray ionization rate\footnote{The rate of electron production in the unit volume by ionization of molecular gas is given by $\zeta n_{\rm{H_2}}$, where the He ionization is taken into account, He/H ratio is equal to 0.09. Our definition of the cosmic ray ionization rate is related to the definition of the parameter used in the UMIST Database for Astrochemistry \citep{McElroy2013}.} is allowed to take values in the range $10^{-17}-10^{-14}$~s$^{-1}$. The highest value of $\zeta$ considered is comparable to the upper estimate of this parameter in molecular clouds in the vicinity of supernova remnants \citep{Vaupre2014,Shingledecker2016}. Increasing cosmic ray ionization rate even more leads to the decline of the molecular hydrogen abundance and the gas becomes almost entirely atomic at $10^{-12}$~s$^{-1}$ \citep{Bayet2011}.

In molecular clouds not far away from sources of ionizing radiation, the ortho-to-para-H$_2$ conversion in collisions of H$_2$ with H$_3^+$ and H$^+$ is expected to be effective -- the time-scale of this process is of the order of 0.1~Myr at the cosmic ray ionization rate 10$^{-16}$~s$^{-1}$ and the gas density $2 \times 10^4$~cm$^{-3}$ \citep{Pagani2013}. As a result, the H$_2$ OPR in molecular clouds at such physical conditions is expected to be close to its steady-state value. At gas temperature $T_{\rm{g}} = 30$~K, the steady-state H$_2$ OPR is equal to 0.03 \citep{Flower2006}. In our simulations, the H$_2$ OPR in the preshock gas has been taken equal to 0.1. The chemistries of ortho- and para-H$_2$ are not distinguished in the model. The ortho-/para-H$_2$ interconversion is considered in reactive collisions H$_2$--H and H$_2$--H$^+$, and via the H$_2$ formation on dust grains.

Following \citet{Godard2019} we define the transition between the shock and the post-shock medium as the point at which most of the molecular radiation (95 per cent) induced by the shock has been emitted. For the sake of simplicity, we assume that the line of sight is perpendicular to the shock front, and the population densities of H$_2$ rotational levels are integrated along the gas flow. Once the column densities $N_{j}$ of rotational levels are obtained, the H$_2$ OPR for a given ortho-H$_2$ level $j$ is calculated as described by \citet{Wilgenbus2000}:

\begin{equation}
\rmn{OPR}_{j} = \rmn{OPR}_{\rmn{LTE},j} \left(T_{\rmn{exc}, j} \right) \frac{N_{j}}{N_{\rmn{LTE}, j} \left(T_{\rmn{exc}, j} \right)},
\label{opr_def_eqn}
\end{equation}

\noindent
where the excitation temperature between levels $j+1$ and $j-1$, $T_{\rmn{exc}, j}$, is given by

\begin{equation}
T_{\rmn{exc}, j} = \frac{E_{j+1} - E_{j-1}}{k_{\rmn{B}} \rmn{ln} \left( N_{j-1} g_{j+1} / N_{j+1} g_{j-1} \right)},
\end{equation}

\noindent
and $N_{\rmn{LTE}, j}$ is the column density obtained by the interpolation between the points $j-1$ and $j+1$ assuming local thermodynamic equilibrium at temperature $T_{\rmn{exc}, j}$. For a range of rotational levels, for example, for levels $2 \leq j \leq 6$, the H$_2$ OPR is calculated using the equation (\ref{opr_def_eqn}) for ortho-H$_2$ levels $j = 3$ and 5, and then averaged. The H$_2$ excitation diagram -- the column densities in rotational levels (divided by the statistical weight $g_{\rmn{v,j}}$ of the level) as a function of the level energy -- may exhibit curvature, indicating that the emitting gas has a range of kinetic temperatures. The error introduced by this curvature is discussed by \citet{Wilgenbus2000}.

\begin{table}
\caption{Parameters of the shock model.}
\label{table:modelparam}
\begin{tabular}{p{4.5cm} p{3cm}}
\hline
Parameter & Value \\
\hline
Preshock gas density, $n_{\rm{H, tot}}$ & $2 \times 10^3-2 \times 10^5$~cm$^{-3}$\\
Shock speed, $u_{\rm{s}}$ & $5-100$~km~s$^{-1}$ \\
Cosmic ray ionization rate, $\zeta$ & $10^{-17} - 10^{-14}$~s$^{-1}$ \\
Initial ortho-to-para-H$_2$ ratio & 0.1 \\ 
Magnetic field strength, $\beta$ & 1 \\
Visual extinction, $A_{V}$ & 10 \\
Micro-turbulence speed, $v_{\rm{turb}}$ & 0.3~km~s$^{-1}$ \\
Grain surface area density & 9.7$\times 10^{-22}$~cm$^{2}$~per~H \\
\hline
\end{tabular}

\medskip
$\beta$ is the dimensionless value of the transverse (to the shock velocity) magnetic field, $\beta = B [\rm{\mu G}] /\left(n_{\rm{H,tot}} [\rm{cm}^{-3}] \right)^{1/2}$.
\end{table}

\section{Results}

\begin{figure}
\includegraphics[width=\columnwidth]{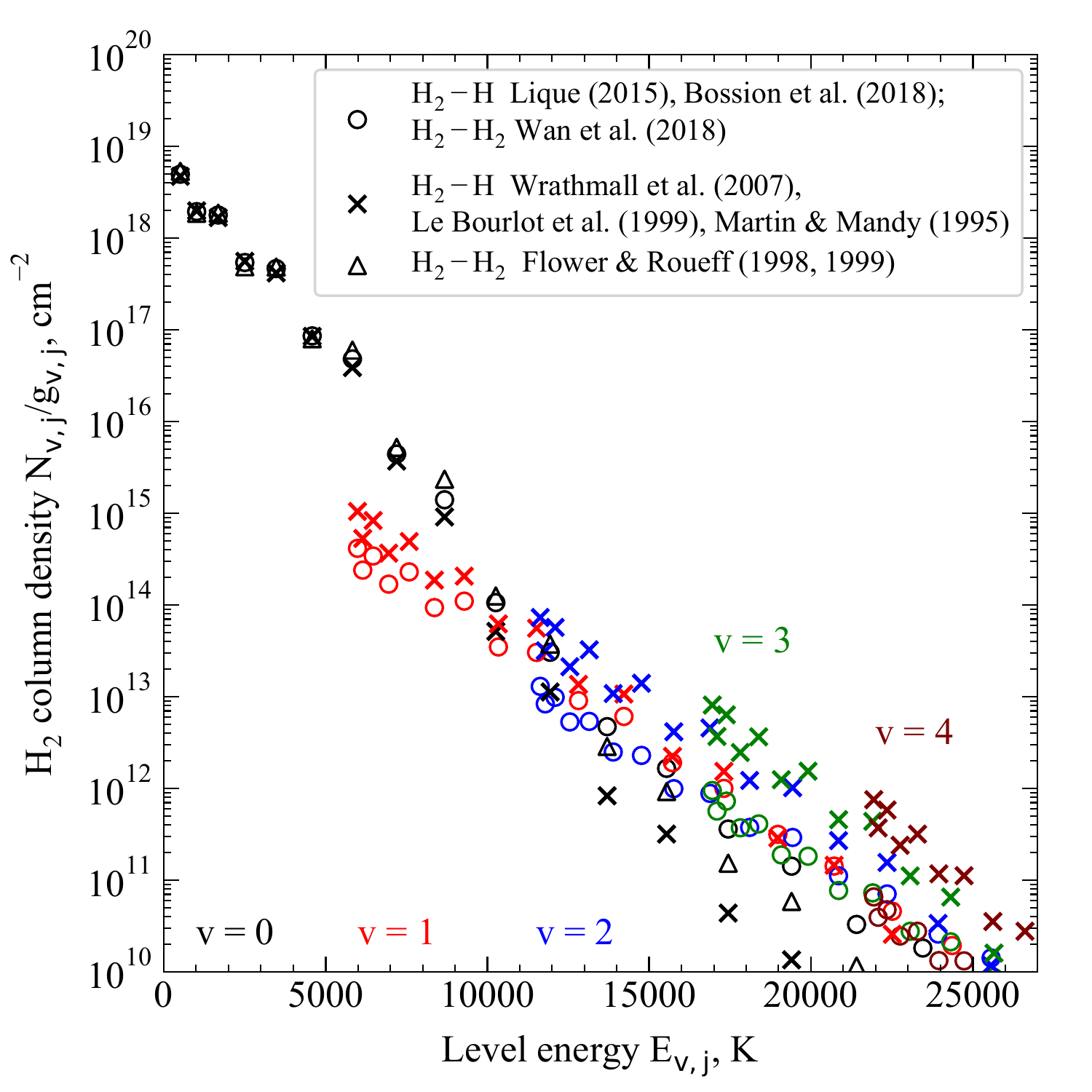}
\caption{The H$_2$ excitation diagram for the C-type shock model using the different data on H$_2$--H and H$_2$--H$_2$ collisions. Parameters of the model are: $n_{\rm{H,tot}} = 2\times 10^4$~cm$^{-3}$, $u_{\rm{s}}=25$~km~s$^{-1}$, $\zeta = 10^{-15}$~s$^{-1}$. {\it{Circles}}: the simulations employ the data for H$_2$--H collisions by \citet{Lique2015,Bossion2018}, data for H$_2$--H$_2$ collisions by \citet{Wan2018} for transitions involving H$_2$ energy levels within the ground vibrational state ($j \leq 31$), and by \citet{Flower1998b,Flower1999} for other transitions. {\it{Crosses}}: the H$_2$--H collisional rate coefficients by \citet{Martin1995,LeBourlot1999,Wrathmall2007} are used. {\it{Triangles}}: the rate coefficients by \citet{Flower1998b,Flower1999} are used for all H$_2$--H$_2$ collisional transitions. In the latter case, only column densities for energy levels of the ground vibrational state of H$_2$ are shown. The vibrational state of the energy level is marked by the color of the data point. For the ground vibrational state, the column densities for energy levels with angular momentum $j \geq 2$ are shown.}
\label{fig_h2_excitation_coll_data}
\end{figure}

\subsection{H$_2$ excitation diagram}
Fig.~\ref{fig_h2_excitation_coll_data} shows the H$_2$ excitation diagram for the C-type shock models using the different data on collisional rate coefficients for H$_2$ molecule, other things being equal. The C-type shock models were calculated with the shock speed $u_{\rm{s}}=25$~km~s$^{-1}$, the preshock gas density $n_{\rm{H,tot}} = 2\times 10^4$~cm$^{-3}$ and the cosmic ray ionization rate $\zeta = 10^{-15}$~s$^{-1}$. The abundance of atomic hydrogen in the preshock gas is $1.7 \times 10^{-3}$, and it is approximately unchanged through the shock region -- the H atom production in chemical reactions is not significant. At H/H$_2$ ratio in question, H$_2$--H collisions are dominant in the excitation of high-lying rotational levels of the ground vibrational state of H$_2$ molecule. Both H$_2$--H and H$_2$--H$_2$ collisions are important in the ro-vibrational excitation of H$_2$.

The populations of low-lying rotational levels of the ground vibrational state of H$_2$ are close to the local thermodynamic equilibrium, and column densities do not depend on the collisional data. But it is no longer valid for the high-lying levels -- it is seen from Fig.~\ref{fig_h2_excitation_coll_data} that populations of the high-lying rotational levels $j \geq 10$ of the ground vibrational state strongly depend on the collisional data for both H$_2$--H and H$_2$--H$_2$ collisions. At high gas temperature $T_{\rm{g}} \simeq 1000$~K, the data by \citet{Lique2015} predict higher rates, by a factor of few, of collisional (de-)excitation of energy levels of the ground vibrational state with high $j$ compared with the data by \citet{Wrathmall2007} (with the inclusion of reactive channels as in \citet{LeBourlot1999}). For energy levels belonging to the vibrationally excited states the situation is reversed -- collisional (de-)excitation rates by \citet{Wrathmall2007} are higher. These differences can be explained by the effect of the reactive hydrogen exchange channels that are not included in the calculations by \citet{Wrathmall2007}, see discussion by \citet{Lique2015}. As a result, the rate coefficients by \citet{Lique2015,Bossion2018} yield higher column densities for rotational levels belonging to the ground vibrational state and lower column densities for the ro-vibrationally excited levels than the data by \citet{Wrathmall2007}. The higher the vibration quantum number of the energy level, the larger the difference between simulations. Column densities of ro-vibrationally excited levels have smaller scatter from a line on the excitation diagram for new rate coefficients. 
%The vibrational quenching is almost as probable as rotational quenching for levels with $v \geq 1$ \citep{Lique2015}.

The data by \citet{Wan2018} on H$_2$--H$_2$ collisions lead to lower (de-)excitation rates of H$_2$ molecule than the data by \citet{Flower1998b,Flower1999} by a factor of about $1.5-2$ for low rotational levels and by a factor of about 10 for levels with $j \geq 10$ at $T_{\rm{g}} \simeq 1000$~K. As a result, the rate of H$_2$ cooling is less effective in the model using the data by \citet{Wan2018}, and the neutral gas temperature at the shock peak is higher by about 10 per cent in this case (the emission in H$_2$ rotational transitions of the ground vibrational state is the dominant cooling mechanism). The H$_2$--H$_2$ collisions do not contribute to the excitation of high-lying rotational levels. The difference between the results of simulations that are based on the data by \citet{Wan2018} and by \citet{Flower1998b,Flower1999} is explained by the difference in the temperature attained by the shocked gas, see Fig.~\ref{fig_h2_excitation_coll_data}.

\begin{figure*}
\includegraphics[width=180mm]{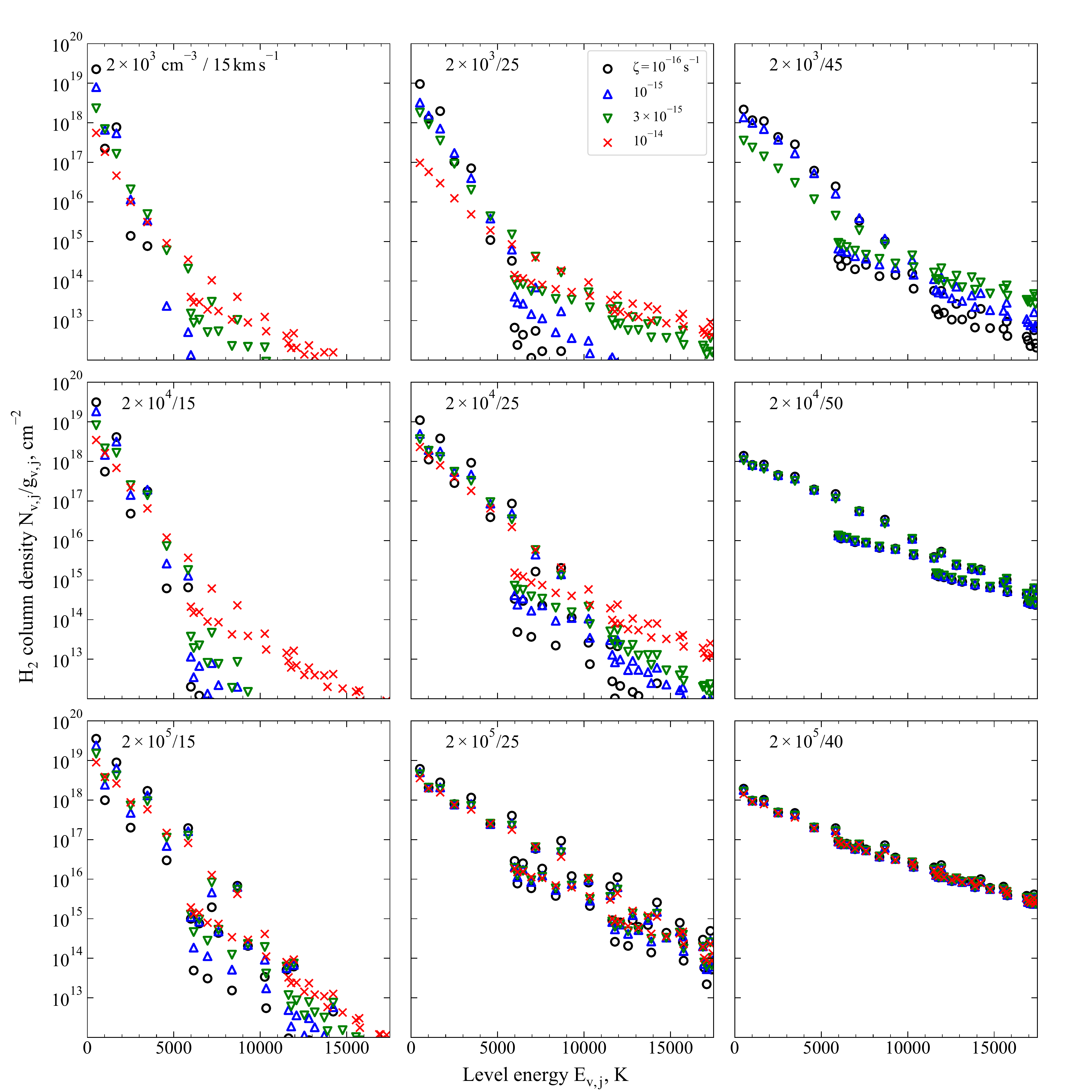}
\caption{The H$_2$ excitation diagrams for C-type shock models with different speeds, preshock gas densities and cosmic ray ionization rates. The plots in one row correspond to the fixed density of the preshock gas, results for gas densities $n_{\rm{H,tot}} = 2 \times 10^3$, $2 \times 10^4$, $2 \times 10^5$~cm$^{-3}$ are shown. The preshock gas density in cm$^{-3}$ and shock speed in km~s$^{-1}$ are indicated in each plot. The highest shock speed for which the results are shown are 45, 50, 40~km~s$^{-1}$ at $2 \times 10^3$, $2 \times 10^4$ and $2 \times 10^5$~cm$^{-3}$, respectively.}
\label{fig_h2_excitation_phys_param}
\end{figure*}

Fig.~\ref{fig_h2_excitation_phys_param} shows excitation diagrams of H$_2$ for shock models computed for a range of preshock gas densities, shock speeds and cosmic ray ionization rates. There is a strong sensitivity of H$_2$ excitation diagram (especially for ro-vibrationally excited levels) to the cosmic ray ionization rate. The width of C-type shock (and, hence, the temperature attained by the gas) depends on the momentum transfer rate between neutral and ionized fluids. At low gas densities ($n_{\rm{H,tot}} \lesssim 10^4$~cm$^{-3}$), there is a strong dependence of the shock width on the ionization fraction of the gas, which is determined by cosmic ray ionization. This effect was discussed by \citet{Wardle1999,Gusdorf2012}. Hence, the cosmic ray ionization affects the excitation of H$_2$ rotational levels in two ways -- it increases the abundance of atomic hydrogen, and enhances the temperature of the shocked gas, see Fig.~\ref{fig_h2_excitation_phys_param}. At $n_{\rm{H,tot}} \gtrsim 10^5$~cm$^{-3}$, the charged dust grains dominate the momentum transfer and heating of the neutrals \citep{Wardle1998,Chapman2006}. Ionization fraction of the gas has small effect on the structure of C-type shock at such densities. However, high cosmic ray fluxes affect hydrogen atom abundance that has significant effect on the H$_2$ excitation diagram at low shock speeds, see Fig.~\ref{fig_h2_excitation_phys_param}. At high shock speeds, the abundance of atomic hydrogen is increased by H$_2$ dissociation, and H$_2$ level populations approach local thermodynamic equilibrium. The effect of cosmic rays diminishes in this case. The total energy emitted by H$_2$ molecule at the given shock speed is almost independent on the cosmic ray ionization rate, see also \citet{Godard2019}. Note, that \cite{Godard2019} conducted a similar study of the H$_2$ excitation in C-type shocks depending on different physical conditions, but in the presence of a strong external UV radiation field.

The non-thermal electrons produced by cosmic rays are an important excitation mechanism of H$_2$ vibrational states (along with the H$_2$ excitation in formation process) at low shock velocities, $u_{\rm{s}} \leq 15$~km~s$^{-1}$, and at moderate cosmic ray ionization rates, $\zeta \leq 10^{-15}$~s$^{-1}$. However, column densities of vibrationally excited levels of H$_2$ are low in this case, see Fig.~\ref{fig_h2_excitation_phys_param}. At higher shock speeds, the gas temperature in the shock is high enough that collisional excitation by gas species dominates the level populations of low vibrational states. The same holds for higher cosmic ray ionization rates. This agrees with the conclusions by \citet{Tine1997}, who pointed out that H$_2$ excitation by non-thermal electrons is important at gas temperatures $T_{\rm{g}} \lesssim 1000$~K. 

The effect of H$_2$--H$^+$ collisions on the column densities of some vibrationally excited levels of H$_2$ is of the order of 10 per cent at low shock speeds and at moderate cosmic ray ionization rates. We note that effect of H$_2$ collisions with ions may be substantial if all abundant ions (which might undergo proton-exchange reactions with H$_2$) will be taken into account. For example, the abundance of H$_3^+$ can be $1-2$ orders of magnitude higher than that of H$^+$.

\begin{figure*}
\includegraphics[width=180mm]{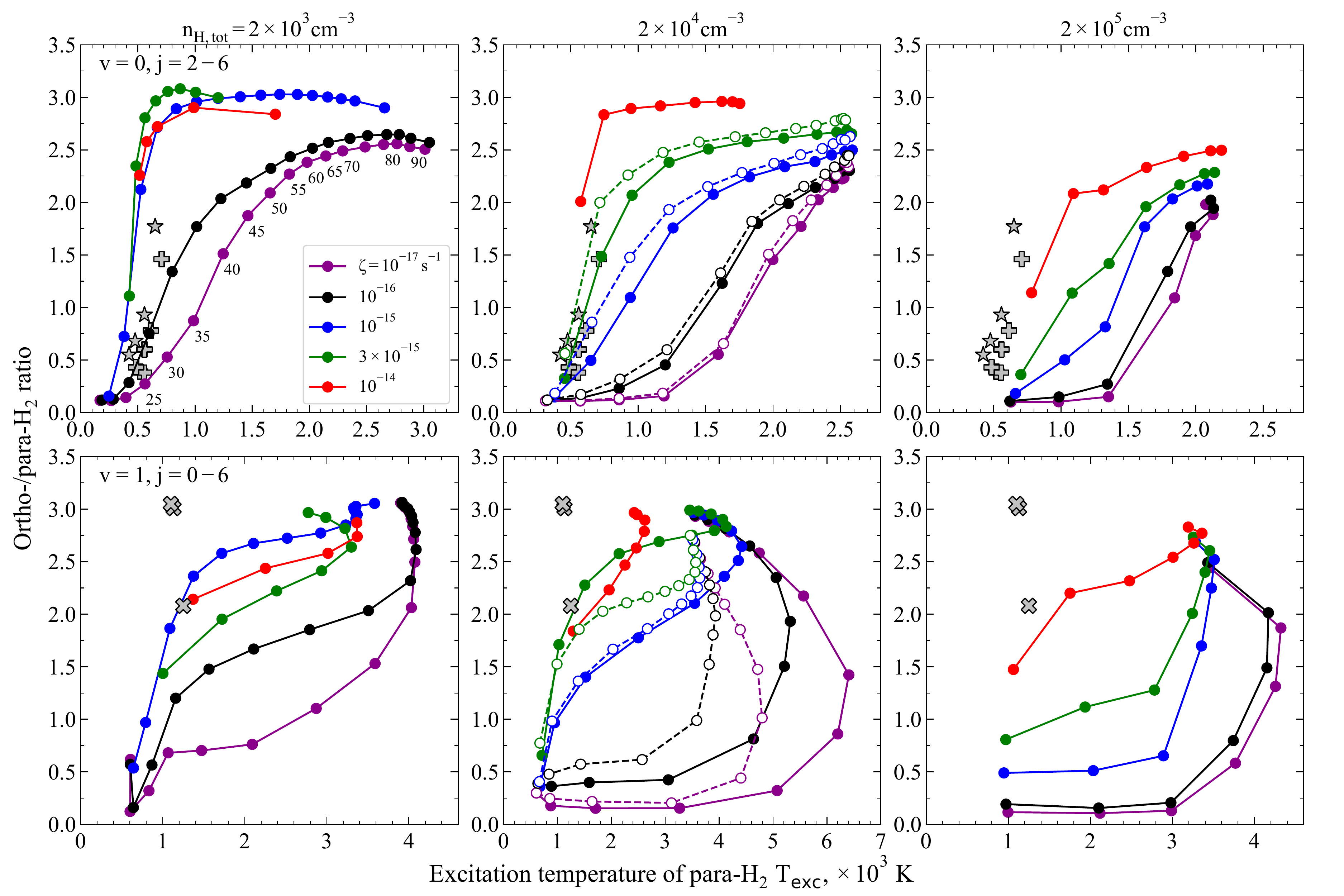}
\caption{The H$_2$ OPR plotted against the mean excitation temperature of para-H$_2$ for a range of shock speeds, cosmic ray ionization rates and different preshock densities. The H$_2$ OPR and the mean excitation temperature of para-H$_2$ are calculated for rotational levels $j = 2-6$ belonging to the ground vibrational state (top panels), for levels $j = 0-6$ belonging to the vibrational state $v = 1$ (bottom). The preshock gas density $n_{\rm{H,tot}}$ is indicated at the top of the panel column. Each point on a given curve represents a different shock speed, starting from 10~km~s$^{-1}$ with an interval 5~km~s$^{-1}$ (the shock speeds are indicated on the upper-left panel for $\zeta$ = 10$^{-17}$~s$^{-1}$ curve). For $n_{\rm{H,tot}} = 2 \times 10^4$~cm$^{-3}$, the simulations based on the data by \citet{Wrathmall2007,LeBourlot1999} are shown (dashed line). The stars on the top panels show the H$_2$ OPR and the excitation temperature for supernova remnants W28, 3C~391, W44, IC~443C (in the order of increasing H$_2$ OPR) that are calculated based on the observational data by \citet{Neufeld2007}. The plus markers show the data on supernova remnants G348.5-0.0, 3C~396, G349.7+0.2, Kes~69, Kes~17, G346.6-0.2 that are calculated based on the observational data by \citet{Hewitt2009}. The filled cross markers on the bottom panels show the data on supernova remnant IC~443 (regions B, C, G). In this case, the H$_2$ OPR and the excitation temperature are estimated based on the observed column densities of energy levels $v=1$, $j = 1-3$ \citep{Shinn2011}.}
\label{fig_h2_ortho_para}
\end{figure*}

\subsection{Para-to-ortho-H$_2$ conversion}
For the shock model presented in Fig.~\ref{fig_h2_excitation_coll_data}, the H$_2$ OPR -- deduced from column densities integrated over the shock width and averaged over the energy levels $2 \leq j \leq 6$ of the ground vibrational state -- is equal to 1.8. The same H$_2$ OPR is for energy levels of the first vibrational state ($0 \leq j \leq 6$). The data by \citet{Wrathmall2007} yield about 10 per cent higher OPR for energy levels of the ground vibrational state and about 5 per cent lower for the first vibrationally excited state. The abundance of atomic hydrogen in the preshock gas is relatively high at the cosmic ray ionization rate in question, and the para-to-ortho-H$_2$ conversion is effective. The H$_2$ OPR reaches the high-temperature limit 3 in the hot shocked gas. However, there is a non-negligible contribution to H$_2$ emission of the initial part of the shock, where the gas temperature is high enough to excite vibrational states, but H$_2$ OPR has not enough time to reach the high-temperature limit, see also \cite{Wilgenbus2000}. The spin conversion process in H$_2$--H collisions is few times slower than the spin conserving process at high gas temperature \citep{Lique2015}. 

The correlation between the H$_2$ OPR and the excitation temperature of H$_2$ energy levels is often considered as a tracer of physical parameters of the shock wave (e.g., \citealt{Wilgenbus2000,Neufeld2006}). Fig.~\ref{fig_h2_ortho_para} shows the mean H$_2$ OPR for rotational levels belonging to the $v = 0$ and 1 vibrational states plotted against the mean excitation temperature of para-H$_2$. The data by \citet{Lique2015} predict lower para-/ortho-H$_2$ interconversion rates than the rates estimated based on \citet{Wrathmall2007,LeBourlot1999}. In particular, for the transitions $j = 2 \to 1$ and $j = 4 \to 3$ within the ground vibrational state, the data by \citet{Lique2015} predict about $2-3$ times lower de-excitation rates than the rate of para-to-ortho-H$_2$ conversion given by \citet{Schulz1965,Schofield1967} at $500-1000$~K. As a result, data by \citet{Lique2015} predict lower H$_2$ OPR for levels of the ground vibrational state, see Fig.~\ref{fig_h2_ortho_para}. New rate coefficients by \citet{Lique2015,Bossion2018} predict higher excitation temperature and higher integrated H$_2$ OPR (at moderate and high shock speeds) for the vibrationally excited energy levels than the analogous simulations with the data by \citet{Wrathmall2007,LeBourlot1999}, see Fig.~\ref{fig_h2_ortho_para}.

Fig.~\ref{fig_h2_ortho_para} shows that the H$_2$ OPR of the shocked gas strongly depends on the cosmic ray ionization rate at moderate shock speeds -- there is a substantial increase of the H$_2$ OPR as the cosmic ray ionization rate increases from 10$^{-17}$ to 10$^{-15}$~s$^{-1}$. On the other hand, the excitation temperature of H$_2$ levels depends both on the shock speed and the cosmic ray ionization rate. The H$_2$ OPR is less than 3 for the ground and vibrationally excited states for most of the shock models. At high cosmic ray ionization rates and at high shock speeds, the integrated H$_2$ OPR may be slightly higher than the local thermodynamic equilibrium value of 3, see Fig.~\ref{fig_h2_ortho_para}. At such physical conditions, the local H$_2$ OPR may attain values up to $3.5-4$ in the initial part of the shock. The reason of H$_2$ OPR values higher than 3 is that level populations are not in local thermodynamic equilibrium. The critical densities (at which the probabilities of collisional and radiative de-excitation are equal) are $n_{\rmn{H_2}} \sim 10^4-10^5$~cm$^{-3}$ for high-lying rotational levels $(v,j) > (0,7)$ (at $T_{\rmn{g}} \simeq 5000$~K). At low gas density, most of the molecules populate low-lying rotational levels even at high gas temperature. There is a high ratio of statistical weights for energy levels with low angular momentum in the limit of high temperature, for example, for levels $j$ = 5 and 4 this ratio is equal to 3.7. On the other hand, if the H atom concentration is relatively high (at high cosmic ray ionization rate or due to dissociation in the high speed shock), the characteristic time of para-to-ortho-H$_2$ conversion is lower than the time needed for the gas temperature reach its peak in the shock, and the H$_2$ OPR attains its local equilibrium value. As the gas cools and becomes dense downstream, the H$_2$ OPR approaches local thermodynamic equilibrium value of 3. We note that para-to-ortho-H$_2$ transitions are $3-5$ times faster than ortho-to-para-H$_2$ transitions \citep{Lique2012}. The effect of high H$_2$ OPR (higher than 3) in low density gas is also discussed by \citet{LeBourlot1999}.

The H$_2$ OPR of ro-vibrationally excited levels may be higher or lower than that of the ground vibrational state depending on shock parameters. The effect of H$_2$--H$^+$ collisions on the H$_2$ OPR and the excitation temperature is negligibly small.

\begin{figure}
\includegraphics[width=\columnwidth]{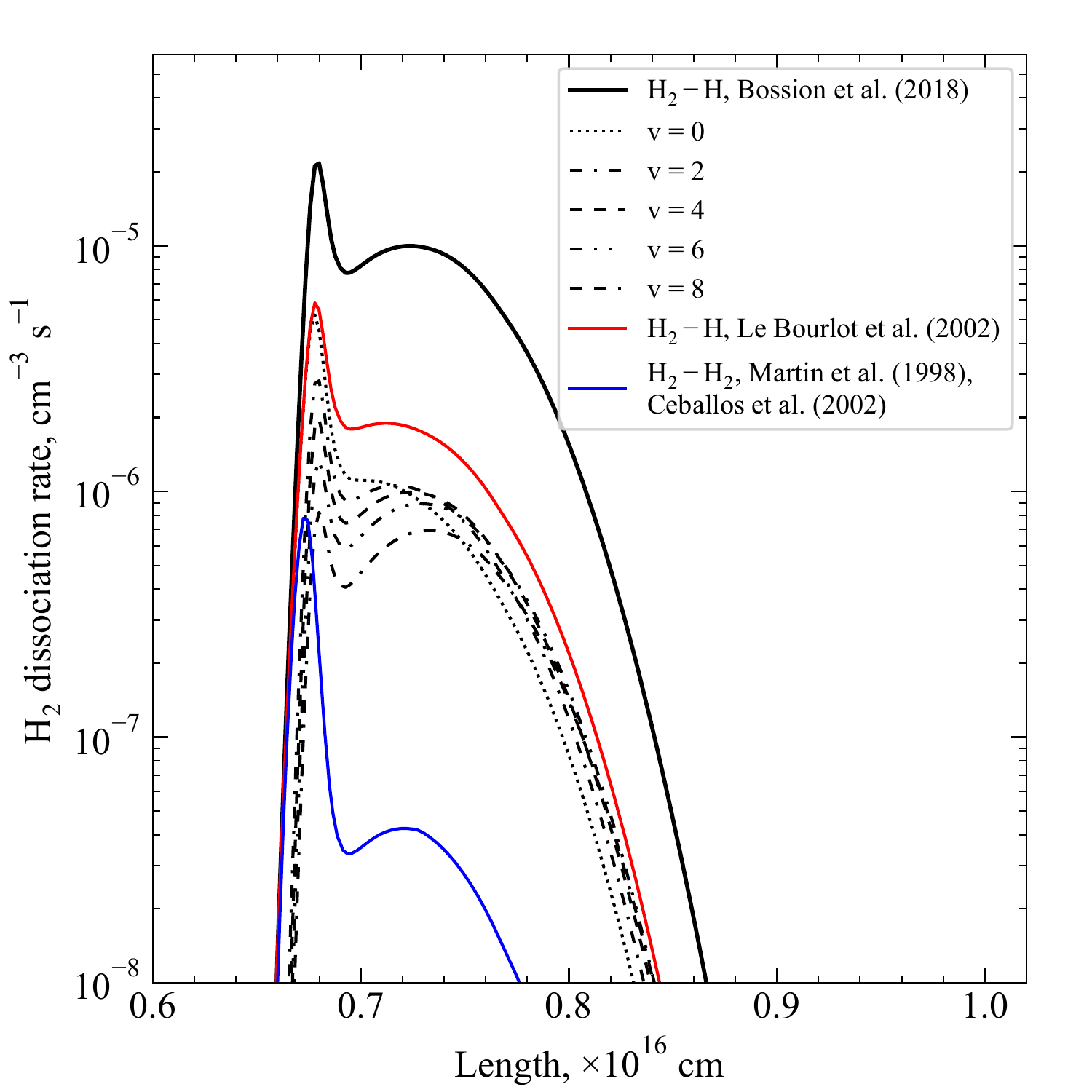}
\caption{The rate of H$_2$ dissociation in the hot shocked gas at the cosmic ray ionization rate of $10^{-16}$~s$^{-1}$, the preshock gas density of $2 \times 10^4$~cm$^{-3}$ and the shock speed of 65~km~s$^{-1}$, the shock speed is close to critical. The contribution of each of the vibrational states $v = 0,2,4,6,8$ to the H$_2$--H dissociation channel is shown. The dissociation rate in H$_2$--H collisions calculated using the formula given by \citet{LeBourlot2002} is also shown.}
\label{fig_h2_diss}
\end{figure}

\subsection{Dissociation of H$_2$ molecules in H$_2$--H collisions}
At high shock speeds, the main route of H$_2$ destruction in the hot gas is H$_2$ dissociation by collisions with H atoms. According to our calculations, the dissociation of H$_2$ in H$_2$--H$_2$ collisions contributes about $1-5$ per cent to the total H$_2$ dissociation rate, dissociation from both the ground and vibrationally excited states being important. The H$_2$ destruction in collisions with ions and electrons is negligibly small at high shock speeds.

At some 'critical' speed $u_{\rm{crit}}$ of the shock wave, there is almost all dissociation of molecular hydrogen. In this case the C-type solution ceases to exist, the shock wave becomes a J-type \citep{LeBourlot2002}. Fig.~\ref{fig_h2_diss} shows the dissociation rate of H$_2$ in H$_2$--H collisions at shock speed just below the critical speed. At high shock speeds, there is a fast rise of neutral gas temperature up to $15~000-20~000$~K at the beginning of the shock as populations of H$_2$ and, hence, the H$_2$ cooling do not respond instantaneously to the changes of physical conditions. As vibrationally excited states of H$_2$ become populated, the H$_2$ cooling is effective and the gas temperature falls. The first sharp peak in the H$_2$ dissociation curve in Fig.~\ref{fig_h2_diss} is explained by the fast rise of the neutral gas temperature. While the second peak -- by the increase of the number densities of atomic hydrogen and H$_2$ molecule on vibrationally excited states, that is lagged behind the temperature peak. The H$_2$ dissociation rate in H$_2$--H collisions depends on the number density of H$_2$ molecule on the vibrationally excited states (see Fig.~\ref{fig_h2_diss}). Thus, the inclusion of collisional excitation of high-lying rotational levels is important in calculations of H$_2$ dissociation rates. \citet{LeBourlot2002} provided an approximate formula for the H$_2$ dissociation rate in collisions with H atoms which is based on the QCT calculations by \citet{Dove1986}. The data by \citet{Bossion2018} predict $3-5$ times higher dissociation rate than an estimate based on \citet{LeBourlot2002}, other things being equal.

At preshock gas density $n_{\rm{H,tot}} = 2 \times 10^3$~cm$^{-3}$ and low cosmic ray ionization rates, our calculations predict $u_{\rm{crit}} \simeq 100$~km~s$^{-1}$ that is about 25 per cent higher than critical speed at this density published by \citet{LeBourlot2002}. At $n_{\rm{H,tot}} = 2 \times 10^4$~cm$^{-3}$ and $2 \times 10^5$~cm$^{-3}$, our simulations provide $u_{\rm{crit}} \simeq 65$~km~s$^{-1}$ and $u_{\rm{crit}} \simeq 40$~km~s$^{-1}$, respectively, that are close to critical speeds found by \citet{LeBourlot2002}. Elevated levels of ionization rate lead to higher degree of ionization of the gas and higher H atom abundance in the preshock gas -- critical velocities are lower in this case, see also discussion by \citet{LeBourlot2002,Melnick2015,Godard2019}. At $\zeta = 10^{-14}$~s$^{-1}$, the maximal shock speed, consistent with the existence of a C-type shock solution, is about 30, 45, 40~km~s$^{-1}$ at $n_{\rm{H,tot}} = 2 \times 10^3$, $2 \times 10^4$, $2 \times 10^5$~cm$^{-3}$, respectively. The dependence of critical speed on ionization rate is stronger at low gas density. Here we consider only solutions where the neutral flow stays supersonic over its entire trajectory. This condition may be broken at low gas densities and at high intensities of ionizing radiation \citep{Godard2019}.

\section{Discussion}
\subsection{Physical conditions in the preshock gas}
The rate of para-to-ortho-H$_2$ conversion in the shock is determined by the abundance of atomic hydrogen. Usually, the gas-phase production of water from atomic oxygen is considered as the main source of H atoms in warm gas at moderate shock speeds (e.g., \citealt{Neufeld2006}). But most of the oxygen is locked up in H$_2$O on icy mantles of dust grains in our model, and this channel is not so effective in H atom formation as considered in previous works. The oxygen atoms presented in the gas-phase are adsorbed on dust grains and become hydrogenated on the grain surface during the chemical evolution of the dark cloud. The gas-phase chemical reactions -- production of water, ammonia, methane -- have noticeable effect on the abundance of atomic hydrogen in the shock only at low cosmic ray ionization rates, $\zeta \sim 10^{-17}$~s$^{-1}$. The H atom abundance in the shocked gas is low in this case, H/H$_2 \sim 10^{-5}-10^{-4}$. This explains low H$_2$ OPR at low cosmic ray ionization rates and at low and moderate shock speeds, see Fig.~\ref{fig_h2_ortho_para}. At $\zeta \gtrsim 10^{-16}$~s$^{-1}$, ionization rate is the main parameter controlling the abundance of atomic hydrogen and, hence, it determines the rate of para-to-ortho-H$_2$ conversion. It explains strong dependence of the H$_2$ OPR on the cosmic ray ionization rate, see Fig.~\ref{fig_h2_ortho_para}. In addition to the elevated fluxes of low-energy cosmic rays, molecular gas near supernova remnants also may be subject to the X-ray emission from the supernova interior and to the UV radiation produced by nearby fast shocks, see also discussion by \citet{Yuan2011}.

The H/H$_2$ ratio in the preshock gas is close to the chemical equilibrium in our model. However, the molecular gas may not have a steady-state H/H$_2$ ratio at the time of passage of a shock wave. First, the molecular cloud may be younger than the characteristic time of H$_2$ formation from atomic gas that is of the order of 1~Myr for a cloud core with gas density $10^4$~cm$^{-3}$ \citep{Goldsmith2007}. In this case cloud core will have H/H$_2$ ratio higher than equilibrium. Secondly, there is a finite response time of the dark cloud chemistry to the increase of the cosmic ray ionization rate. The H/H$_2$ ratio reaches equilibrium value at about $(3-5) \times 10^4$~yr after the change of cosmic ray ionization rate (at $n_{\rm{H,tot}}=2 \times 10^4$~cm$^{-3}$ and using the 'standard' H$_2$ production rate on dust grains). This time-scale does not depend on the ionization rate but is inversely proportional to the rate of H$_2$ formation. Regarding the second item, one can expect that H/H$_2$ ratio is close to its steady-state value for molecular clouds in the vicinity of middle-aged supernova remnants (with ages of about few 10$^4$~yr). 

Other important parameter is the H$_2$ OPR in the preshock gas. The ortho-H$_2$ decay time is about $0.5-1$~Myr at $\zeta = 10^{-17}$~s$^{-1}$ and gas density of $10^4$~cm$^{-3}$, and is in inverse ratio with the cosmic ray ionization rate and the gas density \citep{Pagani2013,Bovino2017,Furuya2019}. In molecular gas around young protostars, where the ionization rate is relatively low, the H$_2$ OPR in the preshock gas may not reach its equilibrium value at the time the protostellar outflow arrives the cloud. For example, \cite{Nisini2010,Giannini2011} analysed the {\it{Spitzer}} Infrared Spectrograph maps of H$_2$ pure rotational lines towards four outflows from Class 0 sources. They estimated the initial H$_2$ OPR in the preshock gas being close to 1. Note, accreting protostars can accelerate cosmic rays, and the ionization rate in the surrounding cloud may be comparable or even higher than due to Galactic cosmic rays \citep{Podio2014,Padovani2016,Gaches2018}. At high cosmic ray ionization rate, $\zeta \simeq 10^{-14}$~s$^{-1}$, the steady-state H$_2$ OPR in the preshock gas is close to 1. In this case, the H$_2$ OPR quickly reaches the high temperature equilibrium value in the shock, and the results are almost independent of the initial H$_2$ OPR.

In our model, cosmic rays are the primary ionization agents and drivers of ion--molecule chemistry. However, at low visual extinctions, $<5$~mag, interstellar UV radiation becomes the dominant source of ionization, and it significantly affects the chemistry of the gas. \citet{Melnick2015}, and more recently \citet{Godard2019}, considered the propagation of low velocity molecular shocks in environments illuminated by a strong external UV radiation field. High levels of UV radiation, as cosmic rays, affect the structure of the C-type shock wave via the enhancement of the photoionization mechanisms and the increase of the dissociation of H$_2$ molecule \citep{Godard2019}. However, the chemical effects of high fluxes of cosmic rays are less pronounced. Even at the highest cosmic ray ionization rates detected in dense clouds in the vicinity of supernova remnants, the gas is molecular (H/H$_2 \sim 0.01$) and photodesorption is not effective enough to evaporate icy mantles of dust grains. However, we show that there is a strong dependence of the H$_2$ excitation on the cosmic ray ionization rate in C-type shocks.

\subsection{Comparison with observations}
\citet{Neufeld2007} carried out spectroscopic mapping observations using the Infrared Spectrograph (IRS) of the {\it{Spitzer Space Telescope}} towards the regions where a supernova remnant interacts with a molecular cloud -- W44 region~E, W28~F, 3C~391, and IC~443~C. These observations led to the detection of the S(0)--S(7) pure rotational lines of molecular hydrogen. Emission in these lines likely originates in molecular gas subject to a slow, nondissociative shock \citep{Neufeld2007}. \citet{Hewitt2009} carried out analogous observations towards another six supernova remnants that show evidence of shocked molecular gas. The calculated H$_2$ OPR and the mean excitation temperature are plotted in Fig.~\ref{fig_h2_ortho_para} for these data. Mean excitation temperatures of para-H$_2$ rotational levels $2 \leq j \leq 6$ lie in the range $400-700$~K that are consistent with the presence of non-dissociative shocks of speed about $10-20$~km~s$^{-1}$. \cite{Tram2018} modelled the bow shock by a statistical distribution of planar shocks, and found that emission by bow shock is generally dominated by low-velocity shocks. 

\cite{Vaupre2014} used molecular line observations to constrain cosmic ray ionization rate in molecular clouds in the vicinity of the W28 supernova remnant. Towards positions located close to the supernova remnant, they found cosmic ray ionisation rates of the order of $10^{-15}$~s$^{-1}$. \citet{Indriolo2010} observed interstellar absorption of H$_3^+$ along sightlines that pass through diffuse gas near the supernova remnant IC~443 and deduced ionization rates of the same order. The H$_2$ OPR of $0.5-1.5$ and excitation temperatures $400-700$~K for low-lying pure rotational levels are consistent with the cosmic ray ionization rate $\zeta \simeq 10^{-15}$~s$^{-1}$ and preshock gas density $n_{\rm{H,tot}} \simeq 10^4$~cm$^{-3}$, see Fig.~\ref{fig_h2_ortho_para}. The observations of H$_2$ excitation must be supplemented with the independent determination of ionization rate or gas density in order to reveal a unique set of physical conditions, see also discussion by \cite{Godard2019}. We note, that \citet{Gusdorf2012} discussed the effect of cosmic ray ionization rate on the populations of rotational levels of H$_2$ in their analysis of the W28~F, but the dependence of H$_2$ OPR on the cosmic ray ionization rate was not considered in their work.

\citet{Shinn2011,Shinn2012} presented near-infrared spectral studies of the shocked H$_2$ gas in the supernova remnant IC~443 using the Infrared Camera aboard the {\it{AKARI}} satellite. \citet{Shinn2012} estimated the H$_2$ OPR in IC~443~B equal to $2.4 \pm 0.4$ and $2.1 \pm 0.3$ for levels ($v$, $j$) = (0, $11-13$) and (1, $1-3$), respectively. For regions C and G in IC~443, the H$_2$~OPR for ro-vibrationally excited levels is close to 3. \citet{Shinn2012} suggested that obtained non-equilibrium OPR for region B probably originates from the reformation of H$_2$ on dust grains behind the dissociative J-type shock. According to our calculations, the H$_2$ OPR for vibrationally excited rotational levels is lower than 3 for most of the models of C-type shocks at $\zeta \leq 10^{-15}$~s$^{-1}$, see Fig.~\ref{fig_h2_ortho_para}. The value obtained for IC~443~B is consistent with non-dissociative shock wave in the medium with high cosmic ray ionization rate. The H$_2$ OPR close to 3 may be result of the emission from dissociative shock wave or the high initial value of H$_2$ OPR. The difficulty in interpretation of observations is that multiple shocks, both non-dissociative and dissociative, might be present along the line of sight (e.g., \citealt{Neufeld2007}).  

The important question that is not discussed in the current study is the shapes of spectral lines of H$_2$ molecule and velocity shifts between lines of para-H$_2$ and ortho-H$_2$. Recently, \citet{Neufeld2019} carried out spectrally resolved observations of pure rotational lines of H$_2$ molecule using the EXES instrument on {\it{SOFIA}} towards the shock region HH7. They reported the detection of systematic velocity shifts between the low-lying rotational lines of ortho-H$_2$ and para-H$_2$ that provides the first definitive evidence for the conversion of para-H$_2$ to ortho-H$_2$ behind shock waves. The shock models that use different data on collisional rate coefficients for H$_2$ molecule predict similar behaviour of OPR for low-lying rotational levels of H$_2$, see Fig.~\ref{fig_h2_ortho_para}. In this case, the effect of new collisional data on the magnitude of the velocity shifts between ortho- and para-H$_2$ rotational lines is expected to be small.

\section{Conclusions}
We study the effect of new H$_2$--H collisional rate coefficients published by \cite{Lique2015,Bossion2018} on the excitation of molecular hydrogen in C-type shocks. The new H$_2$--H collisional data are state of the art and are based on the most accurate H$_3$ potential energy surface. The main results of the paper are summarized below:

\begin{enumerate}
\item 
At high gas temperatures, the data by \citet{Lique2015} predict lower rates, by a factor of $1-3$, of collisional (de-)excitation of H$_2$ rotational levels belonging to vibrationally excited states compared with the data by \citet{Wrathmall2007} with the inclusion of reactive channels. As a result, the rate coefficients for H$_2$--H collisions by \citet{Lique2015,Bossion2018} yield lower column densities of H$_2$ rotational levels of vibrationally excited states than the data by \citet{Wrathmall2007}. 

\item The effect of new collisional data is substantial in the determination of H$_2$ OPR for rotational levels of the first vibrationally excited state. The H$_2$ OPR integrated over the shock length is lower than high-temperature equilibrium value for most of the shock models at $\zeta \lesssim 10^{-15}$~s$^{-1}$. It is valid for rotational levels of both the ground and vibrationally excited states.

\item The para-to-ortho-H$_2$ conversion rate in the shock strongly depends on the cosmic ray ionization rate at moderate shock speeds -- there is a substantial increase of the H$_2$ OPR of the shocked gas as the cosmic ray ionization rate increases from 10$^{-17}$ to 10$^{-15}$~s$^{-1}$. The H$_2$ OPR and the excitation temperature can be used to constrain physical parameters from observations of molecular hydrogen in interstellar shocks.

\item The H$_2$--H collisions are the main H$_2$ dissociation channel at high shock speeds. The data by \citet{Bossion2018} predict several times higher dissociation rate in H$_2$--H collisions than an estimate based on calculations by \citet{Dove1986}. The H$_2$ dissociation takes place from high-lying ro-vibrationally excited H$_2$ levels. The data on collisional excitation of high-lying H$_2$ levels are as important as dissociation rates. 

\item The H$_2$--H$_2$ collisions contribute about $1-5$~per~cent to the total dissociation rate. The dissociation of H$_2$ in H$_2$--H$_2$ collisions from the ground vibrational state and from vibrationally excited states is taken into account.

\item The data by \citet{Wan2018} on H$_2$--H$_2$ collisions lead to lower (de-)excitation rates of H$_2$ molecule than the data by \citet{Flower1998b,Flower1999} by a factor of about $1.5-2$ for low rotational levels. As a result, the rate of H$_2$ cooling is less effective in the model using new data on H$_2$--H$_2$ collisions. 

\item The effect of the excitation of H$_2$ by H$^+$ ions and by non-thermal electrons produced by cosmic rays is non-negligible only at low shock speeds and at moderate cosmic ray ionization rates.
\end{enumerate}

\section*{Acknowledgements}
AVN acknowledges the Russian Science Foundation (RSF) grant 16-12-10225. DB, YS and FL are supported by the 'Programme National Physique et Chimie du Milieu Interstellaire' (PCMI) of CNRS/INSU with INC/INP co-funded by CEA and CNES. YS would like to thank support from the Computing Center of the 'Institut National de Physique Nucl\'eaire' (IN2P3). We are grateful to Dr. Wan for providing the collisional rate coefficients for H$_2$--H$_2$ collisions.

\bibliographystyle{mnras}
%\bibliography{../../interstellar_medium_references,../../chemistry_references}
\bibliography{interstellar_medium_references,chemistry_references}

\appendix
\section{Chemistry}
\label{app_chemistry}

In simulations of chemical evolution of the molecular cloud, species are assumed to be initially in atomic form except for hydrogen, which is assumed to be molecular. A simple zero-dimensional model is considered in these simulations, see \cite{Nesterenok2018}. The calculated chemical composition at $t = 0.5$~Myr is chosen for the shock wave modelling. 

The gas-phase chemical network used here is based on the UMIST Database for Astrochemistry (UDfA), 2012 edition \citep{McElroy2013}. Several updates were done to the chemical network, see \citet{Nesterenok2018}. A set of branching ratios for the reactions involving carbon-chain species provided by \citet{Chabot2013} is used. Photodissociation and photoionisation rates of chemical species in the interstellar radiation field and in the cosmic-ray induced UV flux are updated according to \citet{Heays2017}. The dissociation rate of molecular hydrogen by cosmic-rays was taken according to \citet{Padovani2018}. The photochemistry experiments suggest that the photodissociation of adsorbed molecules is less efficient than the corresponding process in the gas phase \citep{Oberg2016,Murga2019}. Here we use solid/gas photodissociation coefficient ratio equal to 0.1 \citep{Kalvans2018}. Note that the rate of dissociative gas-phase reaction H$_2$--H calculated using the rate coefficients provided by UDfA is more than an order of magnitude higher than the accurate one \citep{Bossion2018}.

The properties and size distribution of dust grains determine the rate of H$_2$ formation and, as a result, the abundance of atomic hydrogen in the cold molecular gas (along with the cosmic ray ionization). A single-size grain model is considered, the grain radius is taken equal to 0.05~\micron \ and the dust--gas mass ratio is equal to 0.01. The corresponding grain surface area is about $10^{-21}$~cm$^{2}$~per~H. The dust temperature in simulations of the chemical evolution of the dark cloud is equal to 10~K, that approximately corresponds to starless cold dense regions at high visual extinctions $A_{V} \sim 10$ \citep{Hocuk2017}. 

Once the H atom adsorbs on the surface of a dust grain, it quickly reacts with one of the icy species (CO, HCO, NO, HNO and others), and, as a result, H$_2$ molecules form through hydrogen abstraction reactions. The rate of H$_2$ formation mostly depends on the adsorption rate of H atoms, which in turn depends on the specific surface area of dust grains and the gas temperature. The effective formation rate of H$_2$ can be expressed as $R = \gamma \, n_{\rm{H, tot}} \, n_{\rm{H}}$, where the rate coefficient $\gamma \simeq (2-3) \times 10^{-17}$~cm$^{3}$~s$^{-1}$ in our simulations. The rate of H$_2$ formation in a dense interstellar gas is poorly known. In diffuse clouds, H$_2$ forms with a rate of $R_{\rm{H_2}} = (3-5) \times 10^{-17}$~cm$^{3}$~s$^{-1}$ \citep{Wakelam2017}. Here, we assume that H$_2$ molecule quickly leaves dust grain after the formation, and that one-third of the binding energy of H$_2$ is deposited statistically as internal excitation of the newly formed H$_2$ molecule \citep{Black1987}. However, the effect of H$_2$ formation on the H$_2$ excitation in C-type shocks is negligibly small.

In our model, the hydrogen atom concentration in the gas is approximately equal to 0.5~cm$^{-3}$ at $\zeta = 10^{-17}$~s$^{-1}$ independent of the gas density. \citet{Goldsmith2005} studied \ion{H}{i} self-absorption in five dark clouds and found central number densities of H atoms in the range $2-6$~cm$^{-3}$ that corresponds to the cosmic ray ionization rate of $\zeta \sim 10^{-16}$~s$^{-1}$.

\label{lastpage}
\end{document}